# Optimal mechanical operation in the vicinity of curved vasculature

Xiaochang Leng[1, *], Xingjian Liu[2, *], Will Toress[3],

Tarek Shazly[2,3]

[1]Institute of Engineering Mechanics,
Nanchang University
Jiangxi, China 330031

[2]State Key Laboratory of Material Processing and Die & Mould Technology,
Huazhong University of Science and Technology
Wuhan, China 430074

[3]College of Engineering and Computing, Biomedical Engineering Program
University of South Carolina
Columbia, SC 29208

[4]College of Engineering and Computing, Department of Mechanical Engineering
University of South Carolina
Columbia, SC 29208

*: both authors contributed equally to this study

**ABSTRACT**

It has been shown that geometrical, structural properties vary along the length of the aortic arch. There is a scarcity of studies focus on the variation in the vessel wall thickness of aortic arch. The central premise of this study is that considering the variation in the vessel wall thickness along the circumference of the aortic arch to be governed by the uniform stress distribution across the vessel wall, meeting the principle of optimal mechanical operation of which the distribution of stress across the vessel wall is assumed to be uniform so as to create a favorable mechanical environment for the mechanosensitive resident vascular cells. Aortic arch was created with image-derived three-dimensional (3D) reconstruction technique. A structure-motivated constitutive model was utilized in the numerical modeling and direct boundary value problem was solved. Stress distribution across the vessel wall under physiological loading condition was predicted in circumferential direction to test the role of the wall thickness. The results showed the variation of the vessel wall thickness in the circumferential direction and uniform distribution of the circumferential stress in the aortic wall which implies a favorable mechanical environment for the resident mechano-sensitive vascular smooth muscle cells. Correlation of geometrical and simulation data support the proposed principle of optimal mechanical operation for the aortic arch.

**Keywords:** Aortic arch arteries; Continuum mechanics; Strain energy function; Elasticity; Finite element modeling

**INTRODUCTION**

The aortic arch, directly connected to the heart, serves as a conduit for blood flow and an elastic reservoir for transforming pulsatile cardiac output into steady circulating flow. Abnormal flow-induced wall shear stress imposed on the vascular endothelial cells usually induces the development of atherosclerosis and aneurysm and results in the bulge of the vessel wall [1, 2]. When the stress exceeds the critical value of vessel wall stress, aortic dissection occurs on the aortic arch [3]. The function of aortic arch is governed by its geometry, composition and mechanical properties of its structural constituents as well as the interplay between them.

Various medical imaging modalities were used to reconstruct the geometry of aortic arch and investigate the hemodynamics of aortic arch under unsteady and pulsatile blood flow condition, assuming the circular cross-section [4]. Fluid-structure interaction study was conducted to investigate the effects of unsteady blood flow on the aortic arch with the assumption of uniform radius and thickness along the longitudinal curvature of aortic arch, modeling the aortic arch as multiple layered composite materials with different Young's modulus for each layer [5]. However, there is a scarcity of study utilizing complicated model for aortic arch.

The principle of optimal mechanical operation was first proposed by Y.C. Fung [6]. This principle is directly supported by the uniform strain hypothesis [7, 8]. Alteration in morphometry was correlated with the strain distribution [9, 10]. Geometrical and structural alterations along the aorta was theoretically investigated [11]. It is validated in the vertebral artery, showing that boundary conditions resulted in structural and compositional remodeling [12-14]. For the longitudinal curvature aortic arch, it still remains a question that if this hypothesis still holds.

The central premise of this study is that thickness along the circumference of the aortic arch meets a principle of optimal mechanical operation ensuring a favorable mechanical environment for mechanosensitive resident cells. The objectives of this study are two-fold. First, the motion tracking technique was applied to reconstruct the geometry of aortic arch. Secondly, a comparative analysis of circumferential stress distribution across the arterial wall was performed between these arterial segments, and the results obtained were interpreted in the framework of a previously elaborated principle of optimal mechanical operation. We speculated that the circumferential variation in the vessel wall thickness is responsible for the optimal mechanical operation. To test this hypothesis, we numerically predict the circumferential stress distribution across the vessel wall thickness at various circumferential locations of the aortic arch with longitudinal curvature.

## MATERIALS AND METHODS

*Vessel isolation*

All tissue handling protocols were approved by the Institutional Animal Care and Use Committee at the University of South Carolina. All hearts were acquired from the Yorkshire Pigs (age 8-12 month, mass 60-70 kg) in the local slaughterhouse, immersed in iced 1% phosphate buffer saline (PBS) solution and transported back to the laboratory. Upon heart arrival, the aortic arch was isolated from the surrounding tissue, washed in phosphate buffered saline (PBS), dissected free of perivascular tissue (Fig. 1).

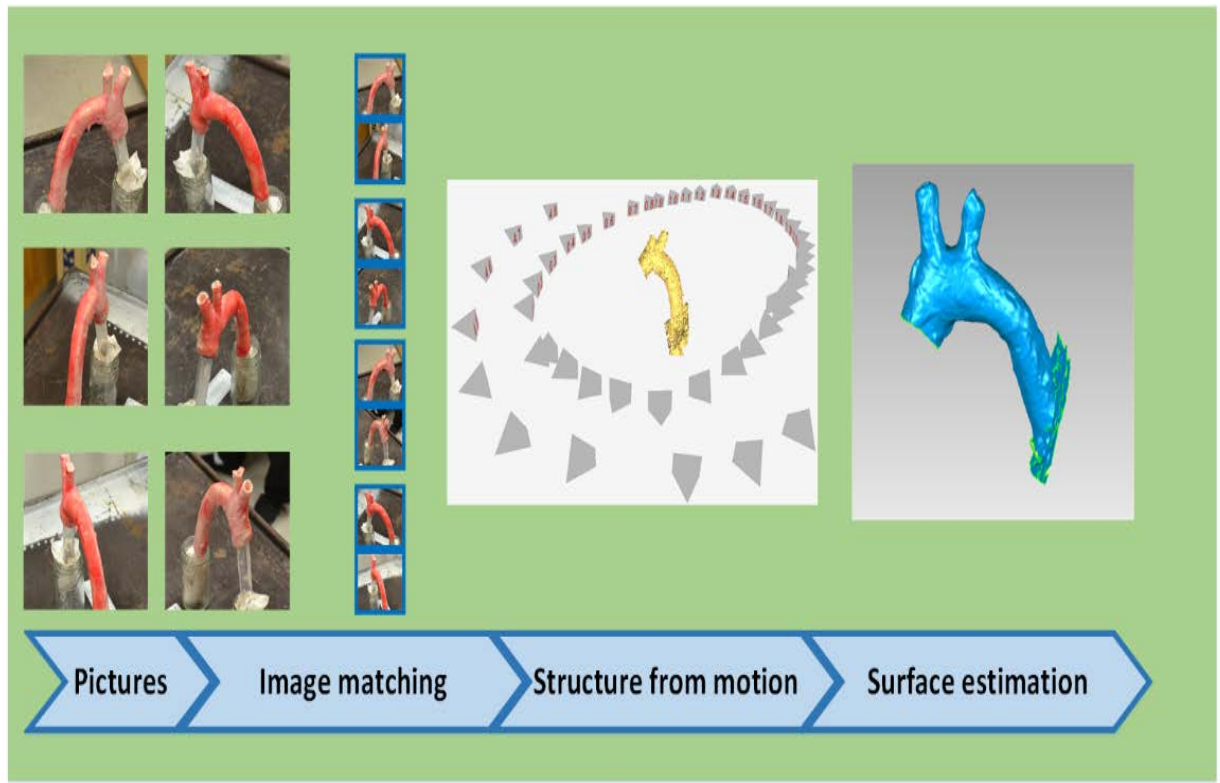

**Figure 1.** Workflow of reconstructing the geometry of aortic arch.

*Lumen formation*

The lumen of the aortic arch was filled with plaster (DAP Plaster of Paris) using 5 ml syringe and wait approximately 30 min for it to be solidified. The plaster was filled a little bit more at the inlet and outlet side for coordinating the tissue and plaster outer surfaces later. In order to increase the brightness contrast so as to improve image quality, blue tissue marking dye was sprayed onto tissue outer surface and multiple images covering the tissue outer surface were acquired with a digital camera. The tissue was then peeled off the plaster and blue tissue marking dye was sprayed onto the plaster surface and multiple images around the plaster were then obtained.

*Geometry reconstruction*

The images of the tissue and plaster outer surfaces obtained from the experiment were processed via image analysis software (Regard3D, MIT) based on OpenMVG library [15]. Firstly, the key points were detected in each image and match them with key points from other images by using AKAZE algorithm [16]. Sparse three-dimensional point cloud of the aortic arch and plaster outer surfaces were reconstructed by using the incremental structure from motion algorithm [17, 18]. It started off with the selected image pair and tries adding other images. Each step was conducted with a bundle adjustment to determine the camera parameters and the 3D positions of the key points [19-21]. To densify the sparse point cloud of the aortic arch and plaster, the CMVS/PMVS tool was used to obtain a dense point cloud [22, 23]. Then the surface data was generated by connecting the dense points. In this process, the Poisson surface reconstruction was applied (Fig. 1) [23]. The aortic wall and plaster were coordinated based on their relative positions [24-26].

*Numerical simulation*

A strain energy function for arterial tissues with distributed collagen fiber orientations (H-G-O Model) was used as follows [27-29],

$$\Psi = \frac{\mu}{2}(I_1 - 3) + \frac{k_1}{k_2}\left[e^{k_2[\kappa \bar{I}_1 + (1-3\kappa)\bar{I}_4 - 1]^2} - 1\right] \tag{1}$$

where $\mu$, $k_1$, and $k_2$ are material constants related to the mechanical properties of elastin and collagen fibers. $\kappa$ is the constant characterizing the distribution of the collagen fibers within two families of fibers (Fig. 2).

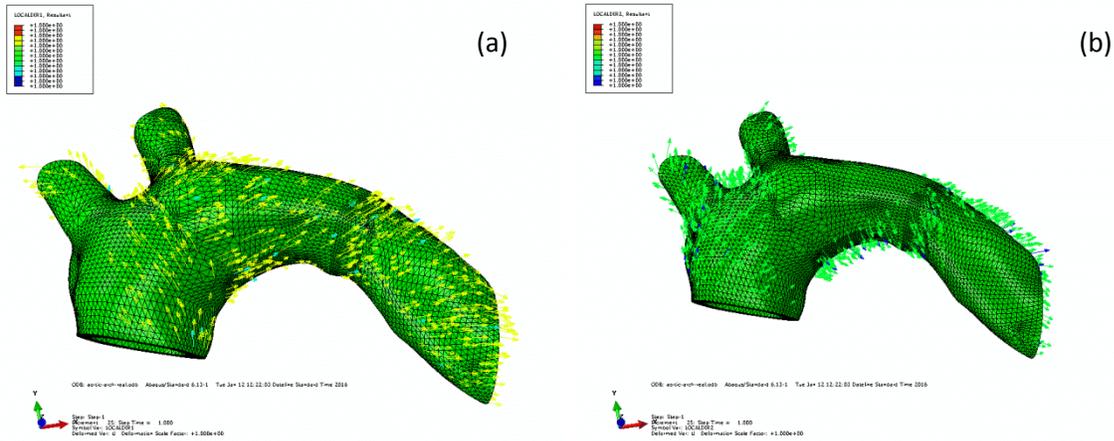

**Figure 2.** Distribution of collagen fibers. (a) 1st and (b) 2nd family of collagen fibers in the aortic arch.

The aortic arch was subjected to the internal pressure of 100 mmHg (Fig. 3a). Material constants of the two fiber family constitutive model were taken to characterize the mechanical property of aorta. The three-dimensional finite element analysis was based on 8-node brick elements. During the computation the top and bottom faces of the aortic arch were fixed (Fig. 3b). The mean arterial pressure was chosen to be P = 100 mmHg. The identified material constants of the strain energy function, are $\mu$ = 3.82 kPa, $k_1$ = 996.6 kPa, $k_2$ = 524.6, $k$ = 0.226. The stress distributions in both the circumferential direction across the vessel wall were predicted.

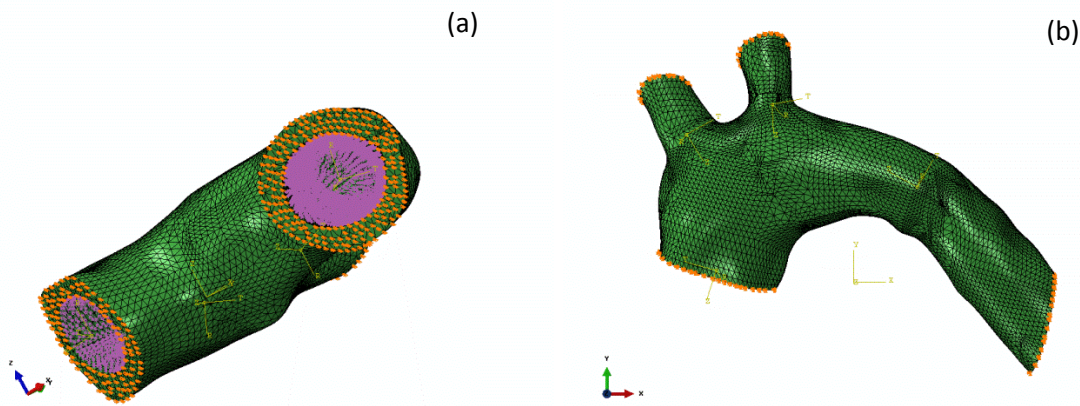

**Figure 3.** Boundary conditions of aortic arch in the modeling. (a) internal pressure; (b) fixed locations during the experiment.

**RESULTS AND DISCUSSION**

The aim of this study was to evaluate the variation of vessel wall thickness of aortic arch, to predict the circumferential stress distribution of vessel wall of aortic arch under physiological loading conditions, and validate the hypothesis of optimal mechanical operation. We expect the longitudinal curvature on circumferential distribution of vessel wall thickness will be retained among other species.

The geometry reconstruction approach utilized in this study for aortic arch is based on the analysis of motion technique (Fig. 1). Vessel wall thicknesses along the longitudinal axis at different cross-sections were measured and compared. The mean values and standard deviations were calculated for the vessel (Table 1). It showed that the thickness decreased from proximal to distal location of the aortic arch (Fig. 4). Vessel wall thicknesses along the circumference were measured and it showed non-uniform distribution of vessel wall thickness along the circumference in one section of the aortic arch (Fig. 5).

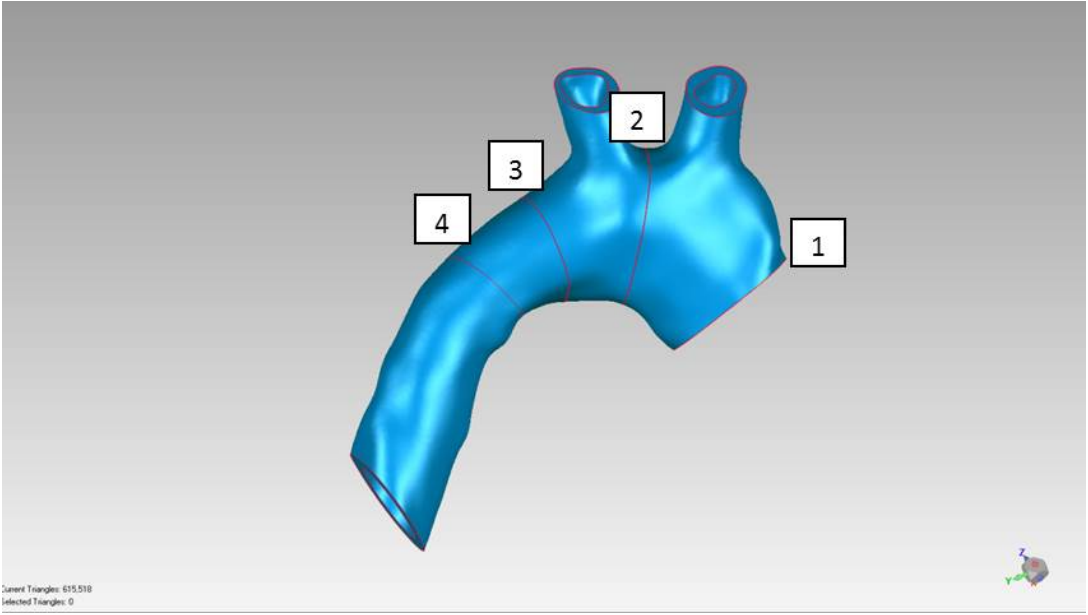

**Figure 4.** Vessel wall thicknesses along the longitudinal axis at different cross-sections.

**Table 1.** Thickness values of different sections along the longitudinal axis of aortic arch.

| Section | Mean [mm] | STD [mm] |
|---------|-----------|----------|
| 1 | 3.05 | 0.58 |
| 2 | 3.47 | 0.81 |
| 3 | 2.37 | 0.38 |
| 4 | 2.27 | 0.44 |

Structure-motivated constitutive model was implemented in the numerical algorithm to predict the circumferential wall stress distribution in each arterial segment under prescribed loading and boundary conditions. The findings showed that the circumferential stress was nearly uniform across the vessel wall given variation in the vessel wall thickness along the aortic arch with longitudinal curvature (Fig. 6). Circumferential stress distribution across the vessel wall thickness at different cross-sections showed uniform magnitude (Fig. 7). These results validated the hypothesis of optimal mechanical operation in the aortic arch.

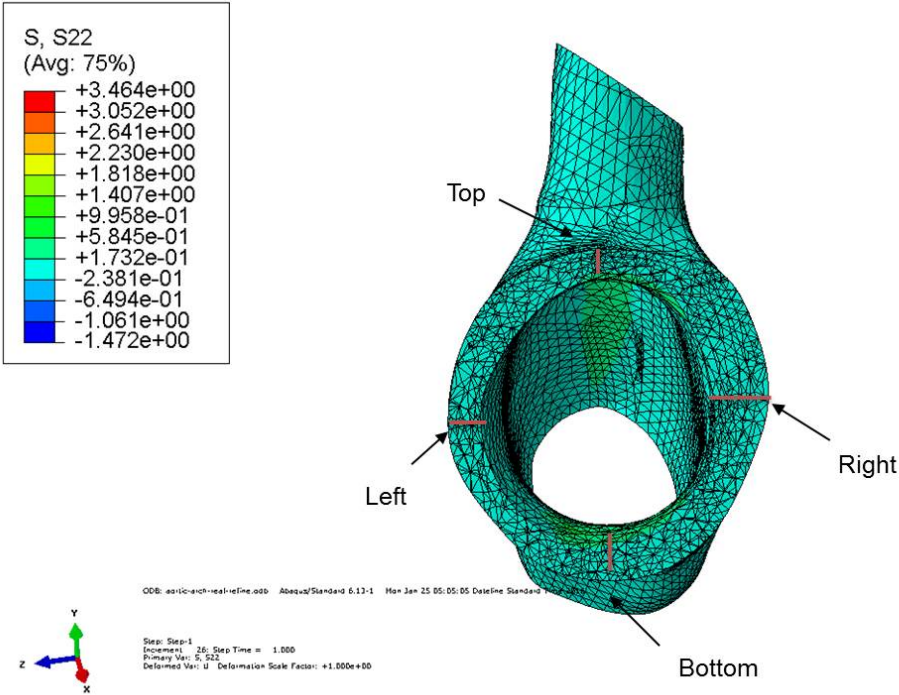

**Figure 5.** Distribution of vessel wall thickness along the circumference in one section of the aortic arch.

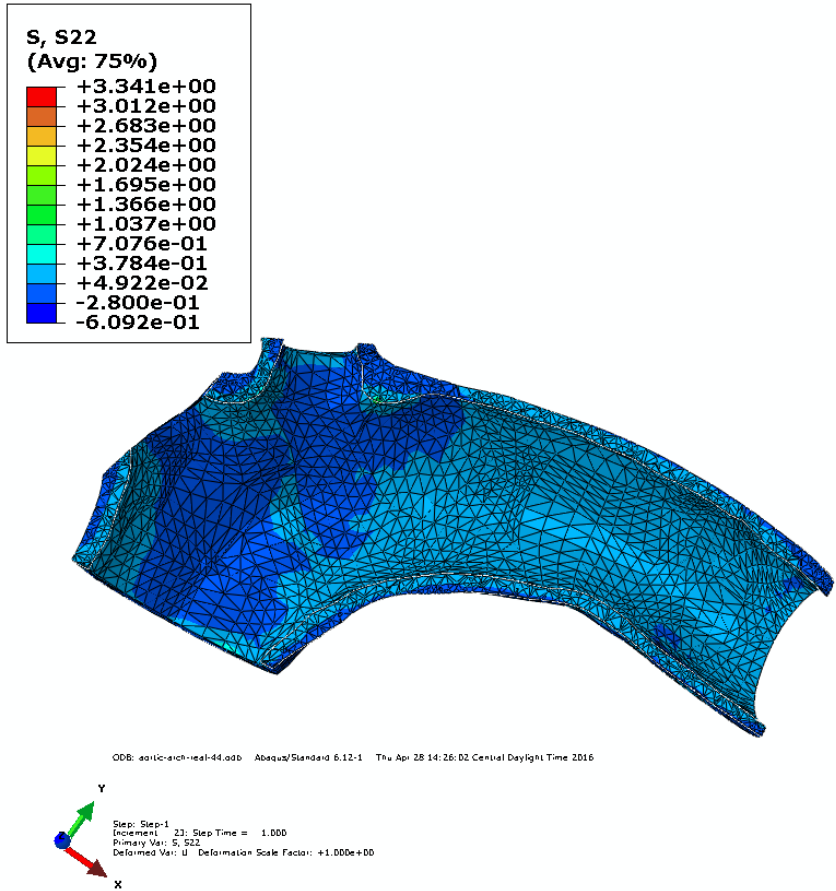

**Figure 6.** Contour of circumferential stress distribution along the aortic arch with longitudinal curvature.

Geometry of aortic arch manifested as arterial branching and curvature leads to highly complex fluid mechanical environment. Nevertheless, vasomotor control of the aortic arch plays an important role in maintaining adequate perfusion to the head and neck while SMC contractile properties are essential in maintaining tissue homeostasis and is an important subject matter for an active vascular mechanics investigation [30].

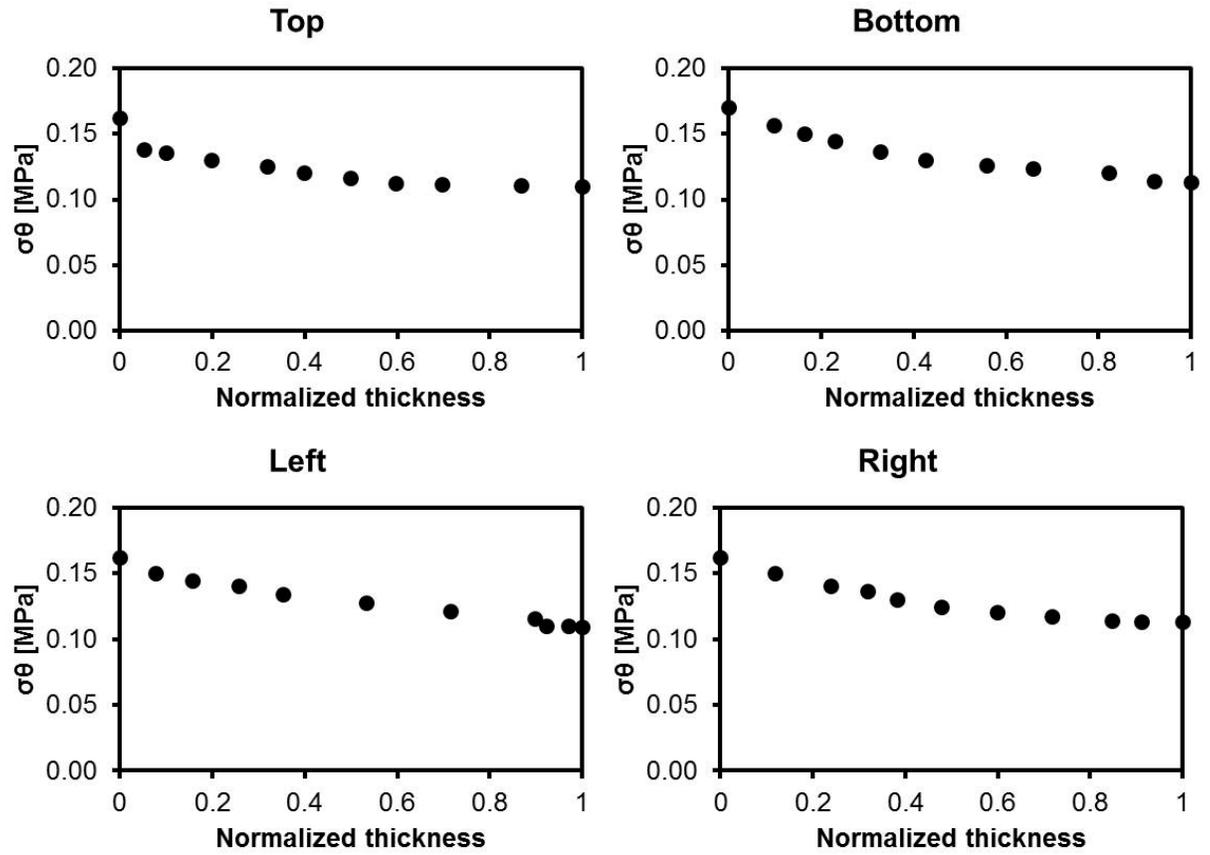

**Figure 7.** Circumferential stress distribution across the vessel wall thickness at different cross-sections.

**CONCLUSION**

In this study, we evaluated the variation of vessel wall thickness of aortic arch at different locations and validate the hypothesis of optimal mechanical operation using numerical simulation. The results showed the variation of the vessel wall thickness in the circumferential direction and uniform distribution of the circumferential stress in the aortic wall which implies a favorable mechanical environment for the resident mechano-sensitive vascular smooth muscle cells. Correlation of geometrical and simulation data support the proposed principle of optimal mechanical operation for the aortic arch.